\documentclass[useAMS,usenatbib,usegraphicx,psfig]{mn2e}

\title[]{The hybrid, coronal lines nova V5588 Sgr (2011 N.2) and its six repeating secondary maxima}
\author[]{U. Munari$^{1}$, A. Henden$^{2}$, D.P.K. Banerjee$^{3}$, N.M. Ashok$^{3}$, G.L. Righetti$^{4}$
\newauthor S. Dallaporta$^{4}$ and G. Cetrulo$^{4}$\\
$^{1}$INAF Astronomical Observatory of Padova, 36012 Asiago (VI), Italy\\
$^{2}$AAVSO, 49 Bay State Rd. Cambridge, MA 02138, USA\\
$^{2}$Astronomy and Astrophysics Division, Physical Research Laboratory,
  Navrangapura, Ahmedabad - 380 009, Gujarat, India\\
$^{4}$ANS Collaboration, c/o Osservatorio Astronomico, via dell'Osservatorio
8, 36012 Asiago (VI), Italy\\}

\begin{document}

\maketitle

\label{firstpage}

\begin{abstract}

The outburst of Nova Sgr 2011 N.2 (=V5588 Sgr) was followed with optical and
near-IR photometric and spectroscopic observations for 3.5 years, beginning
shortly before the maximum.  V5588 Sgr is located close to Galactic center,
suffering from $E_{B-V}$=1.56 ($\pm$0.1) extinction.  The primary maximum
was reached at $V$=12.37 on UT 2011 April 2.5 ($\pm$0.2), and the underlying
smooth decline was moderately fast with $t^V_2$=38 and $t^V_3$=77 days.  On
top of an otherwise normal decline, six self-similar, fast evolving and
bright secondary maxima (SdM) appeared in succession.  Only very few other
novae have presented so clear secondary maxima.  Both the primary maximum
and all SdM occurred at later times with increasing wavelengths, by amounts
in agreement with expectations from fireball expansions.  The radiative
energy released during SdM declined following an exponential pattern, while
the breadth of individual SdM and the time interval between them widened. 
Emission lines remained sharp (FWHM$\sim$1000 km/s) throughout the whole
nova evolution, with the exception of a broad pedestal with a trapezoidal
shape ($\Delta$vel=3600 km/sec at the top and 4500 km/sec at the bottom)
which was only seen during the advanced decline from SdM maxima and was
absent in between SdM.  V5588 Sgr at maximum light displayed a typical
FeII-class spectrum which did not evolve into a nebular stage.  About 10
days into the decline from primary maximum, a typical high-ionization
He/N-class spectrum appeared and remained visible simultaneously with the
FeII-class spectrum, qualifying V5588 Sgr as a rare {\em hybrid} nova. 
While the FeII-class spectrum faded into oblivion, the He/N-class spectrum
developed strong [FeX] coronal lines.

\end{abstract}


\section{Introduction}

V5588 Sgr was discovered at unfiltered 11.7 mag on 2011 Mar 27.832 UT as PNV
J18102135$-$2305306 = Nova Sgr 2011 N.2 by Nishiyama and Kabashima (2011).
Spectroscopic confirmation was obtained on Mar.  28.725 UT by Arai et al.
(2011) who noted prominent emission lines of H$\alpha$ (FWHM=900 km/s),
H$\beta$ and Fe II (multiplets 42, 48, 49) on a highly reddened continuum.
Very red colors were evident in the first photometric observations
obtained on Mar 28.670 UT by Kiyota (2011) and on Mar 28.788 UT
by Maehara (2011), characterized by $B$$-$$V$$\sim$1.7.

The peculiar nature of V5588 Sgr soon started to emerge when Munari et al.
(2011a) reported about a bright and fast evolving secondary maximum reaching
peak brightness on April 25.0 UT, and when Munari et al.  (2011b) observed
a further secondary maximum peaking on May 22.0 UT.  Their optical and IR
spectra showed emission lines having two components, whose relative
intensity greatly changed at the time of the secondary maxima, one narrow with
FWHM=1050 km/s and one broad with full-width-at-zero-intensity (FWZI) of
4700 km/sec.  The intensity ratio between the broader and the narrow
component was larger in HeI than in HI lines.  Rudy et al.  (2011) wrote
about IR spectroscopic observations for April 28 and they too noted the
two-component structure of emission lines, which was also present in the IR
spectroscopic observations for April 26, 28 and May 4 described by Banerjee
and Ashok (2011).  The latter also reported a weak line seen at 2.0894
microns, which was earlier tentatively identified as a coronal line due to
[MnXIV] in the few instances where it has been seen in novae spectra (in
nova V1974 Cyg by Wagner and Depoy 1996; in RS Oph by Banerjee et al.  2009).

Radio observations of V5588 Sgr were first carried out with E-VLA on Apr
21.5, Apr 30.3 and May 1.3 UT by Krauss et al.  (2011a).  These
observations, soon before and after the April 25.0 UT secondary maximum,
failed to detect the nova, which was instead radio-bright when observed on
May 14.5 and 15.5 UT (a week before the May 22 secondary maximum) by Krauss
et al.  (2011b).  According to Krauss et al.  (2011c) the nova was again radio
quiet on June 2.2, radio loud on June 15.4, and once again radio quiet on July
27.1.  Krauss et al.  (2011c) noted how the observed fast rises in
radio emission could not be explained in the framework of a simple expanding
isothermal sphere as modelled in many other novae, and that non-detections
could be taken as upper limits to the thermal flux from the nova ejecta which
indicate that V5588 Sgr is quite distant ($\geq$9 kpc) and likely associated
with the Galactic central regions.

In this paper we present our photometric and spectroscopic monitoring of the
evolution of V5588 Sgr, that tightly covers the first 200 days from
pre-maximum to well into the advanced decline when conjunction with the Sun
prevented further observations.  The later decline was followed by sparser
photometric observation extending up to mid 2014.

\begin{table*}
 \centering
\caption{$B$$V$$R_{\rm C}$$I_{\rm C}$ photometry of V5588 Sgr (the long
table is published in its entirety in the electronic edition of this
journal. A portion is shown here for guidance regarding its form and
content).}
\includegraphics[width=17.5cm]{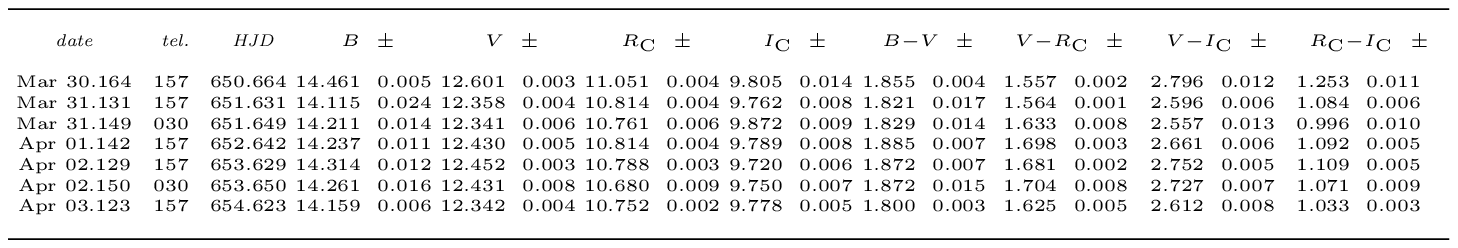}
\end{table*}

\begin{figure}
\includegraphics[width=84mm]{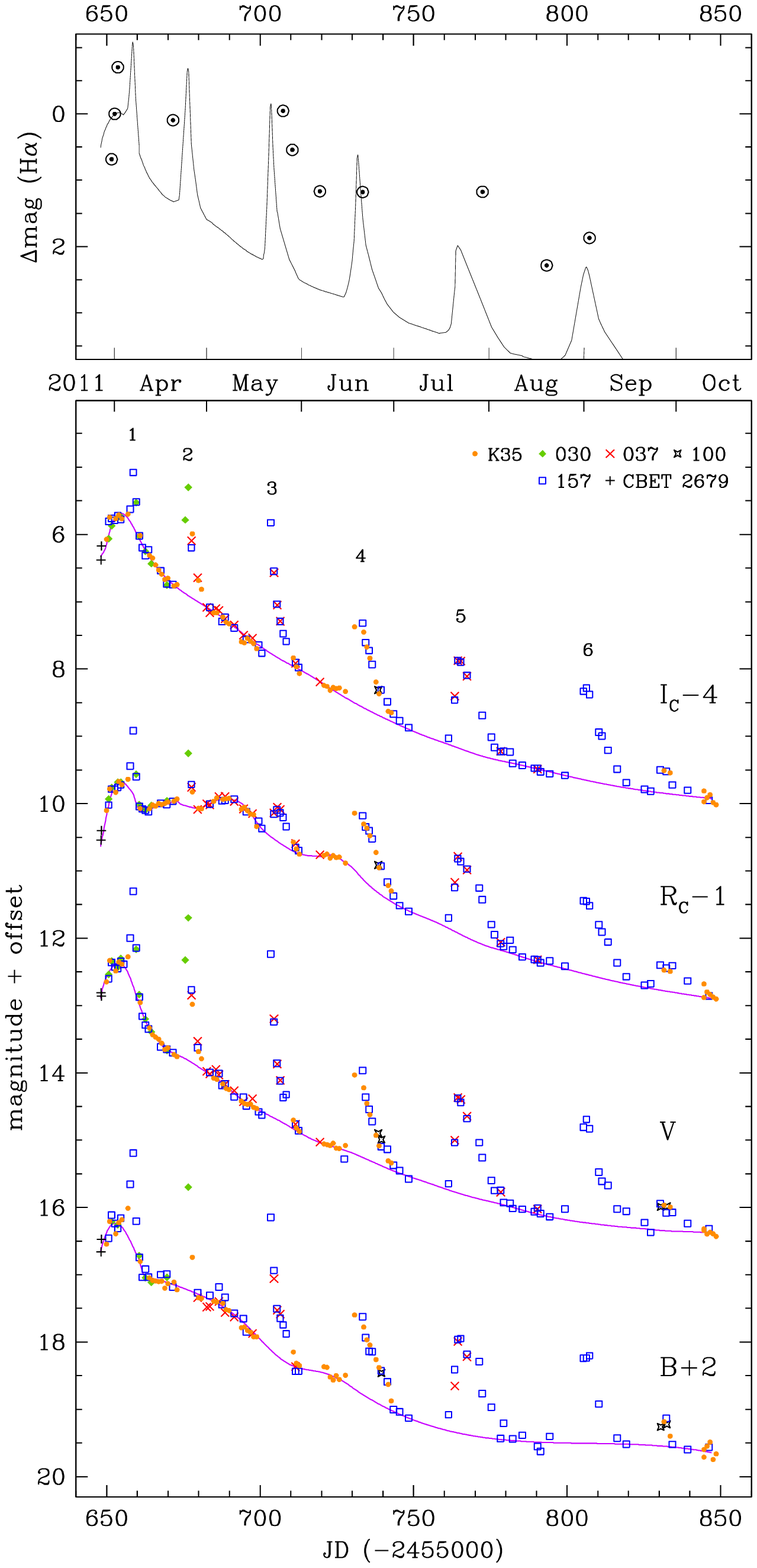}
\caption{{\em Lower panel:} $B$$V$$R_{\rm C}$$I_{\rm C}$ photometric
evolution of V5588 Sgr during 2011 (the different telescopes are identified
in the legend at upper right).  The solid lines are hand-drawn to guide the
eye in following the normal nova decline away from secondary maxima. The
largely different shape of the $R_{\rm C}$ lightcurve is due to the
extremely strong H$\alpha$ emission line that largely dominated the flux
recorded through the band-pass. {\em Upper panel:} evolution of the
H$\alpha$ emission line integrated flux. The ordinate scale is the same of
the lower panel and the zero point is set to 1.080$\times$10$^{-10}$ erg
cm$^{-2}$ sec$^{-1}$ corresponding to the flux for day $-$0.4 spectrum (2011
Apr 02; cf Table~2), the one closest to photometric maximum. For comparison,
the interpolated $V$-band lightcurve from the lower panel is superimposed.}
\end{figure}

\section{Observations}

\subsection{Optical photometric observations}

$B$$V$$R_{\rm C}$$I_{\rm C}$ optical photometry was obtained with ($a$) ANS
Collaboration telescopes N.  30, 37, 100 and 157, and ($b$) AAVSO.net
telescopes K35 and T61.  The same local photometric sequence, spanning a
wide color range and carefully calibrated against Landolt (2009) equatorial
standards, was used at all telescopes and observing epochs, ensuing a high
consistency among different data sets.  The $B$$V$$R_{\rm C}$$I_{\rm C}$
photometry of the nova is given in Table~1, where the quoted uncertainties
are the total error budget, which combines the measurement error on the
variable with the error associated to the transformation from the local to
the standard photometric system (as defined by the photometric comparison
sequence around the nova linked to the Landolt's standards).

The operation of ANS Collaboration telescopes is described in detail by
Munari et al.  (2012) and Munari \& Moretti (2012).  They are all located in
Italy.  The median values of the total error budget of their measurements
reported in Table~1 are: $\sigma$($B$)= 0.016, $\sigma$($V$)= 0.009,
$\sigma$($R_{\rm C}$)= 0.004, $\sigma$($I_{\rm C}$)= 0.007,
$\sigma$($B$$-$$V$)= 0.016, $\sigma$($V$$-$$R_{\rm C}$)= 0.009, and
$\sigma$($V$$-$$I_{\rm C}$)= 0.011.  All measurements on the program nova
were carried out with aperture photometry, the long focal length of the
telescopes and the absence of nearby contaminating stars not requiring to
revert to PSF-fitting.  Concerning ANS Collaboration telescopes, colors and
magnitudes are obtained separately during the reduction process, and are not
derived one from the other.

Both AAVSO.net telescopes K35 and T61 are robotically operated from AAVSO
Headquarters in Cambridge (MA, USA).  K35 (located in Weed NM, USA) was used
during the initial monitoring of the nova up to the end of June 2011, when
the arrival of the monsoon season prevented further observations, which
briefly resumed in October 2011 before the solar conjunction. T61 (located
at Mt. John, New Zealand) was used at later epochs, when longer integrations
were necessary to catch the ever fainter nova. The data reduction of
AAVSO.net data provides $V$ and colors, obtained via aperture photometry
for K35 observations, and PSF fitting for T61.

\begin{table}
 \centering
\caption{Log of the optical spectroscopic observations.}
\includegraphics[width=84mm]{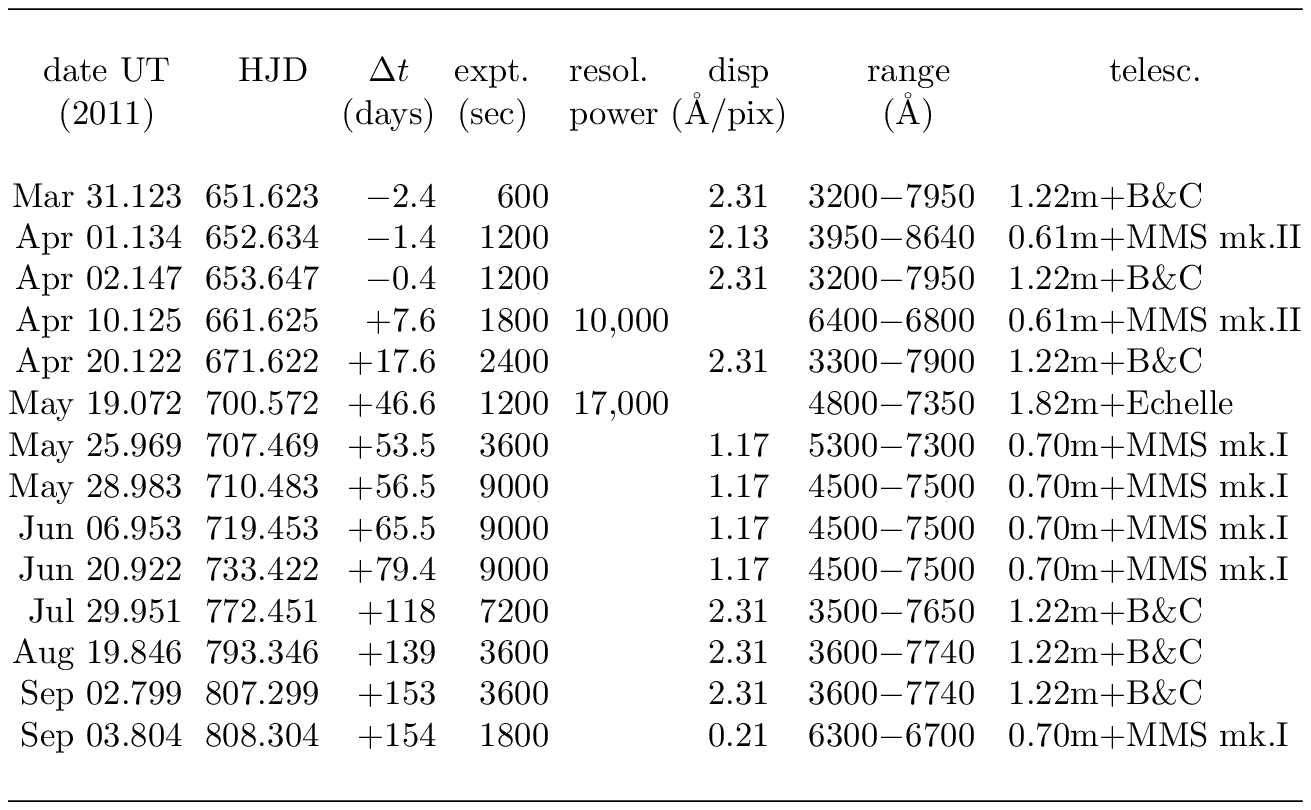}
\end{table}

\begin{figure}
\includegraphics[width=84mm]{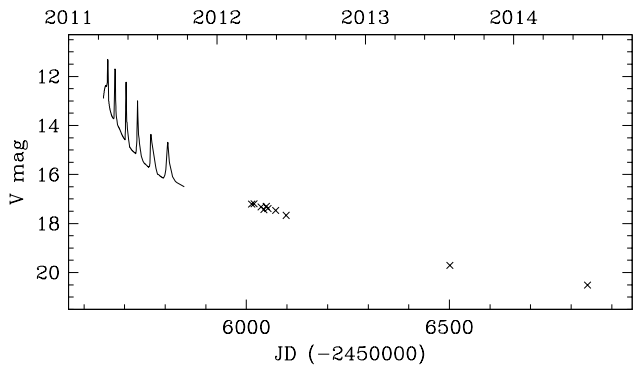}
\caption{$V$-band evolution of V5588 Sgr during the advanced decline.
The interpolated $V$-band lightcurve from Figure~1 is superimposed for comparison.}
\end{figure}

\subsection{Optical spectroscopy}

Optical spectroscopy of V5588 Sgr has been obtained with three telescopes.
A log of the observations is given in Table~2.

ANS Collaboration 0.70m telescope located in Polse di Cougnes (Udine, Italy)
and operated by GAPC Foundation is equipped with a mark.I Multi Mode
Spectrograph, and obtained medium-resolution spectra with a front
illuminated Apogee ALTA U9000 CCD camera (3056$\times$3056 array, 12$\mu$m
pixel, KAF9000 sensor).  ANS Collaboration 0.61m telescope operated by
Schiaparelli Observatory in Varese and equipped with a mark.II Multi Mode
Spectrograph obtained low-resolution and Echelle spectra with a SBIG ST10XME
CCD camera (2192$\times$1472 array, 6.8 $\mu$m pixel, KAF-3200ME chip with
micro-lenses to boost the quantum efficiency).  The optical and mechanical
design, operation and performances of Multi Mode Spectrographs from
Astrolight Instruments in use within ANS Collaboration are described by
Munari and Valisa (2014).

Low resolution spectroscopy of V5588 Sgr was obtained also with the 1.22m
telescope + B\&C spectrograph operated in Asiago by the Department of
Physics and Astronomy of the University of Padova.  The CCD camera is a
ANDOR iDus DU440A with a back-illuminated E2V 42-10 sensor, 2048$\times$512
array of 13.5 $\mu$m pixels.  It is highly efficient in the blue down to the
atmospheric cut-off around 3200 \AA, and it is normally not used long ward of
8000 \AA\ for the fringing affecting the sensor.

The spectroscopic observations at all three telescopes were obtained in
long-slit mode, with the slit rotated to the parallactic angle.  All
observations have been flux calibrated, and the same spectrophotometric
standards have been adopted at all telescopes.  All data have been similarly
reduced within IRAF, carefully involving all steps connected with correction
for bias, dark and flat, sky subtraction, wavelength and flux calibration.

\begin{table}
 \centering
\caption{Near-IR photometry of V5588 Sgr.}
\begin{tabular}{@{}lrrrrrr@{}}
\hline
Date     &  JD  -    &     $J$         &         $H$          &        $K$  \\
(2011)   &  2455000  &               &                    &          \\
\hline
 Apr 26	&	678.387  	&	8.19 $\pm$ 0.04	&	7.59 $\pm$ 0.05	&	6.75 $\pm$ 0.04	\\
 May 04	&	686.473  	&	8.90 $\pm$ 0.02	&	8.73 $\pm$ 0.03	&	7.88 $\pm$ 0.09	\\
 May 07	&	689.442	    	&	9.01 $\pm$ 0.03	&	8.92 $\pm$ 0.04	&	7.97 $\pm$ 0.12	\\
 May 08	&	690.412   	&	9.06 $\pm$ 0.03	&	8.99 $\pm$ 0.03	&	8.11 $\pm$ 0.15	\\
 May 18	&	700.457    	&	9.58 $\pm$ 0.02	&	9.39 $\pm$ 0.05	&	8.59 $\pm$ 0.12	\\
 May 23	&	705.385  	&	9.04 $\pm$ 0.02	&	8.69 $\pm$ 0.02	&	7.95 $\pm$ 0.07	\\
 May 26	&	708.453	    	&	9.40 $\pm$ 0.03	&	9.18 $\pm$ 0.05	&	8.23 $\pm$ 0.12	\\
 June 11&	724.280  	&      10.06 $\pm$ 0.04 &	9.90 $\pm$ 0.05	&	-	\\

\hline
\end{tabular}
\end{table}

\begin{table}
 \centering
\caption{Log of the near-IR spectroscopic observations.}
\begin{tabular}{@{}lccrrrrr@{}}
\hline
Date     &  JD       & $\Delta t$ &\multicolumn{3}{c}{Exp. time (sec)}\\
(2011)   &(-2455000) & (days)     & $J$ & $H$ & $K$  \\
\hline
Apr 26	&	678.463	&	 +24.5 &200	&	200	&	200	\\
May 04	&	686.440	&	 +32.4 &120	&	120	&	120	\\
May 05	&	687.441	&	 +33.4 &120	&	120	&	120	\\
May 06	&	688.458	&	 +34.5 &120	&	150	&	160	\\
May 18	&	700.406	&	 +46.4 &120	&	150	&	160	\\
May 23	&	705.319	&	 +51.3 &150	&	150	&	150	\\
May 24	&	706.365	&	 +52.4 &150	&	150	&	150	\\
May 25	&	707.361	&	 +53.4 &150	&	150	&	150	\\
May 26	&	708.340	&	 +54.3 &120	&	180	&	180	\\
\hline
\end{tabular}
\end{table}

\subsection{Near-infrared observations}

The near-IR photometric and spectroscopic observations of V5588 Sgr were
carried out from 1.2m Mt.  Abu telescope, operated by Physical Research
Laboratory (India).  The $JHK$ spectra were obtained at similar dispersions
of $\sim$ 9.5 Angstrom/pixel in each of the $J, H, K$ bands using the
Near-Infrared Imager/Spectrometer which uses a 256 X 256 HgCdTe NICMOS3
array.  A set of two spectra were taken with the object dithered to two
positions along the slit which were subtracted from each other to eliminate
the sky contribution and the detector dark counts.  The spectra were then
extracted using IRAF and wavelength calibration was done using a combination
of OH sky lines and telluric lines that register with the stellar spectra.
Following the standard procedure, the object spectra were then ratioed with
the spectra of a comparison star (SAO 186061; spectral type B9V) observed at
similar airmass as the object and from whose spectra the Hydrogen Paschen
and Brackett absorption lines had been extrapolated out.  The ratioed
spectra were then multiplied by a blackbody curve at the effective
temperature of the comparison star to yield the final spectra.  Photometry
in the $JHK$ bands was done in photometric sky conditions using the imaging
mode of the NICMOS3 array.  Several frames, in five dithered positions
offset typically by 20 arcsec, were obtained of both the nova and a selected
standard star in each of the $J, H, K$ filters.  Near-IR $JHK$ magnitudes
were then derived using IRAF tasks and following the regular procedure
followed by us for photometric reduction (e.g.  Banerjee $\&$ Ashok 2002,
Naik et al.  2010). The log of the near-IR photometric and spectroscopic
observations are given in Tables 3 and 4.

\begin{figure}
\includegraphics[width=84mm]{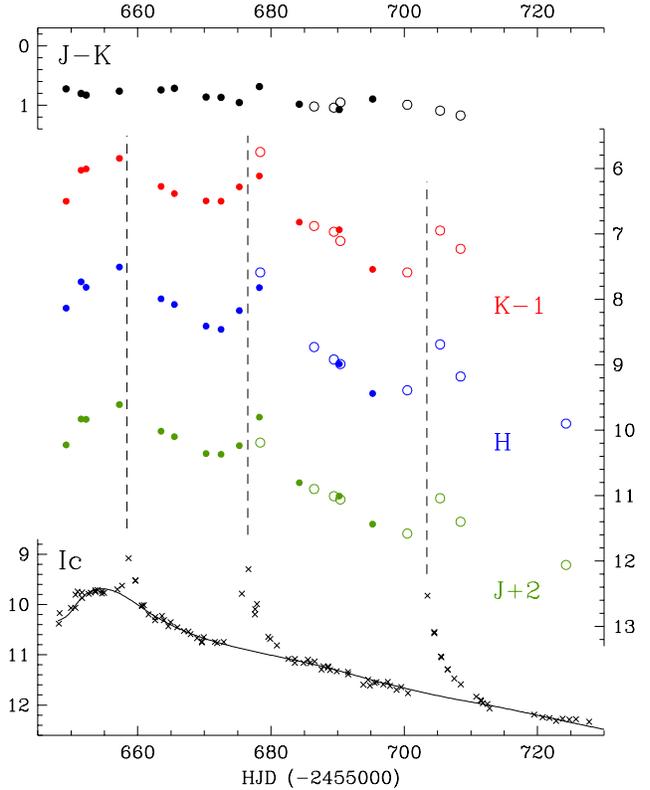}
\caption{Near-IR $J$$H$$K$ photometric evolution of V5588 Sgr. Open circles:
our data from Table~3. Dots: data from Kanata Observatory 1.5m telescope.
The $I_{\rm C}$ lightcurve is reproduced from Figure~1 for comparison with
optical data.}
\end{figure}

\section{Photometric evolution}

The photometric evolution of V5588 Sgr in the $B$$V$$R_{\rm C}$$I_{\rm C}$
bands during 2011 is presented in Figure~1. The 2012-2014 portion of the
lightcurve is displayed in Figure~2, while the evolution in the near-IR
$JHK$ bands is displayed in Figure~3. Figure~4 identifies the remnant in the
surrounding crowded field and Figure~5 provides a zoomed view onto the
secondary maxima and the color evolution.

The photometric evolution of V5588 Sgr is unlike that of other novae: on
top of otherwise smooth and normal-looking maximum and decline phases, six
bright secondary maxima appeared.  These secondary maxima were fast evolving
and followed a clear pattern: their duration and the time interval between
them were increasing with successive maxima.

Secondary maxima appearing on top the lightcurve of otherwise smoothly
evolving novae (thus not to be confused with the rebrightenings that some
slow novae display around maximum, like V723 Cas or V1548 Aql, or around the
transition from optically thick to optically thin conditions, like V1494 Aql
or V2468 Cyg) are quite rare. Some novae (eg. Nova Aql 1994a, Nova Cyg
2006, Nova Cyg 2008b; Venturini et al. 2004, Munari et al. 2008a, Munari et
al. 2011c), have displayed just one such event over their recorded
photometric history.  Strope et al. (2010) in their review of AAVSO
lightcurves of a hundred novae, termed J-type (from "jitter") the novae
showing multiple secondary maxima. Among the lightcurves of J-type novae
they presented, the one closer to V5588 Sgr is - by far - Nova Sgr 2003
(=V4745 Sgr), that displayed five bright secondary maxima but had a
completely different spectroscopic evolution (see below).

We first discuss the underlying smooth photometric evolution of V5588 Sgr as
if belonging to an otherwise normal nova, and then turn our attention to the
secondary maxima.

\subsection{Optical}

The underlying smooth photometric evolution of V5588 Sgr is highlighted in
Figures~1 and 5 by the continuous lines drawn from simple cubic spline
interpolations. The time ($\pm$0.2 days) and brightness of the {\em
normal} maximum (as opposed to the subsequent six secondary maxima, labelled
1 to 6 in Figure~1) as determined from these interpolations are:

\begin{eqnarray}
t_{max}^{B}  &=& 2455653.5     ~~~~B_{max} \,\,= 14.178 \\
t_{max}^{V}  &=& 2455654.0     ~~~~V_{max} \,\,\,= 12.368 \nonumber \\
t_{max}^{Rc} &=& 2455654.3    ~~~~Rc_{max} = 10.654 \nonumber \\
t_{max}^{Ic} &=& 2455654.7    ~~~~Ic_{max} \,\,=  ~9.689 \nonumber
\end{eqnarray}
\noindent

The epoch of maximum occurred at later times for longer wavelength bands,
the maximum in $I_{\rm C}$ coming $\sim$1.2 days after that in the $B$ band.
This delay is similar to the 1.0 days observed in Nova Aql 2009 (V1722 Aql)
by Munari et al.  (2010) who noted how it closely followed the expectations
for the initial fireball expansion phase. At maximum the nova appeared faint
and highly reddened, which contributed to limited interest among observers
and inhibited monitoring programs for e.g. in the X rays or UV.  A wider
interest arose only after the first secondary maxima were reported, but at
that time the nova had already dropped in brightness.

The times (in days) taken by V5588 Sgr to decline by 2 and 3 magnitudes in
the $B$ and $V$ bands are
\begin{eqnarray}
t^B_2 = 50&\phantom{xxxxxx}&t^B_3 = 90  \\
t^V_2 = 38&\phantom{xxxxxx}&t^V_3 = 77  \nonumber
\end{eqnarray}
They are in the normal proportion among themselves (cf Munari et al. 2008a, Eq. 2).

An interesting feature of the early photometric evolution is the single
pulsation-like oscillation that occurred right before maximum (highlighted
in Figure~5).  The single cycle had an amplitude of about $\sim$0.2
mag and was completed in $\sim$4 days.  A similar feature was observed also
in V2615 Oph (Nova 2007) at the time of maximum brightness, with the cycle
having an amplitude of $\sim$1.5 mag and being completed in $\sim$8 days
(Munari et al.  2008b).  In both novae the $B$$-$$V$ color did not change
much in phase with the observed cycle, while $V$$-$$I_{\rm C}$ varied in a
way reminiscent of pulsation (bluer at maximum, redder at minimum). Such
a feature could be related to the pre-maximum halts observed in some novae
(e.g. Hounsell et al. 2010). Theoretical investigations are beginning to
explore in some greater detail the early lightcurve of novae, and
short-lived features similar to the pulsation-like oscillation we observed
in V5588 Sgr are emerging in the computations, as in those of Hillman et al.
(2014).

The $R_{\rm C}$-band lightcurve of V5588 Sgr in Figure~1 looks quite
different from that in the other bands. This is due to the presence of an
extremely strong H$\alpha$ in emission. While other remission lines
counted for a minimal fraction of the flux recorded over the $B$, $V$, and
$I_{\rm C}$ bands, the H$\alpha$ dominated the flux recorded through the
$R_{\rm C}$ band, from 30\% on our first spectrum (day $-$2.4) to 79\% on
the last (Sep 2).

The near-IR photometric evolution of V5588 Sgr is presented in Figure~3
where our data from Table~3 is supplemented with public data from Kanata
Observatory 1.5m
telescope\footnote{http://kanatatmp.g.hatena.ne.jp/kanataobslog/20110514}.
In the near-IR V5588 Sgr behaved similarly to the optical: a smooth
underlying decline on which are superimposed the secondary maxima.  Over the
limited period of time covered by Figure~3, the $J$$-$$K$ color slowly and
smoothly increased by only a small amount, indicating that no warm dust
condensed in the ejecta.  The secondary maxima were not accompanied by
significant $J$$-$$K$ color changes.  Even if the time of primary maximum is
not well covered by $JHK$ data, nonetheless Figure~3 suggests that in the
near-IR it occurred later than in the optical, in accordance with
expectations from an expanding fireball (e.g.  Seaquist \& Bode 2008). 
An important feature displayed by Figure~3 is how the secondary maxima
occurred about two days later in $JHK$ than at optical wavelengths.  Such a
shift and its amount suggest that broad-band emission during SdM were
dominated by expanding photospheres.

\subsection{Reddening and distance}

van den Bergh and Younger (1987) derived a mean intrinsic color
$(B-V)_\circ$=$+$0.23 $\pm$0.06 for novae at maximum, and
$(B-V)_\circ$=$-$0.02 $\pm$0.04 for novae at $t_2$.  The photometric
evolution in Figure~1 and the data in Table~1 show that V5588 Sgr was
measured at $(B-V)$=$+$1.81 at maximum, and $(B-V)$=$+$1.52 at $t_2$.  The
corresponding reddenings are $E_{B-V}$=1.58 and 1.54, for a mean value
$E_{B-V}$=1.56 ($\pm$0.1), which will be adopted in the rest of this paper. 
The corresponding extinction would be $A_V$=5.16 mag following the reddening
relations calibrated by Fiorucci and Munari (2003).  The high reddening
affecting V5588 Sgr is confirmed by the OI emission line ratio in the
infrared observations of Rudy et al.  (2011), and by the flux ratio of
Paschen to Brackett emission lines in our near-IR spectra.

Both the rate of decline and the observed magnitude 15 days past maximum are
popular methods of estimating the distance to a nova.

The relation $M_{\rm max}\,=\,\alpha_n\,\log\, t_n \, + \, \beta_n$ as most
recently calibrated by Downes \& Duerbeck (2000) provides a distance to
V5588 Sgr of 7.8 kpc for $t^V_2$, and 7.5 kpc for $t^V_3$.  The specific
stretched S-shaped curve first suggested in analytic form by Capaccioli et
al.  (1989) gives a distance of 7.1 kpc in the revised form proposed by
Downes \& Duerbeck (2000).  Buscombe \& de Vaucouleurs (1955) suggested that
all novae have the same absolute magnitude 15 d after maximum light.  Its
most recent calibration by Downes \& Duerbeck (2000) returns 8.1 kpc as the
distance to V5588 Sgr.  The straight average of these four determinations is
an absolute magnitude M$_V$=$-$7.2 and a distance of 7.6~kpc, that we
will adopt in this paper.

Such a large distance is consistent with the upper limit of negative radio
detections discussed by Krauss et al.  (2011c).  Considering the galactic
coordinates of the nova $l$=7.84 and $b$=$-$1.88, it seems likely that V5588
Sgr is associated with the Galactic center and the inner Bulge.

\subsection{Astrometric position and progenitor/remnant}

We have derived an accurate astrometric position for V5588 Sgr on K35 images
(3.5m focal length) around maximum brightness. The average position from 8
$R_{\rm C}$ images obtained between 2011-03-30 and 2011-04-02 is (equinox
2000):
\begin{equation}
\alpha=18^h~10^m~21^s.339~~~~~~~~~\delta=-23^\circ~05^\prime~30^{\prime\prime} .09
\end{equation}
with an uncertainty of 40 milli-arcsec on both. Figure~4 provides a deep
finding chart for the remnant.

\begin{figure}
\includegraphics[width=84mm]{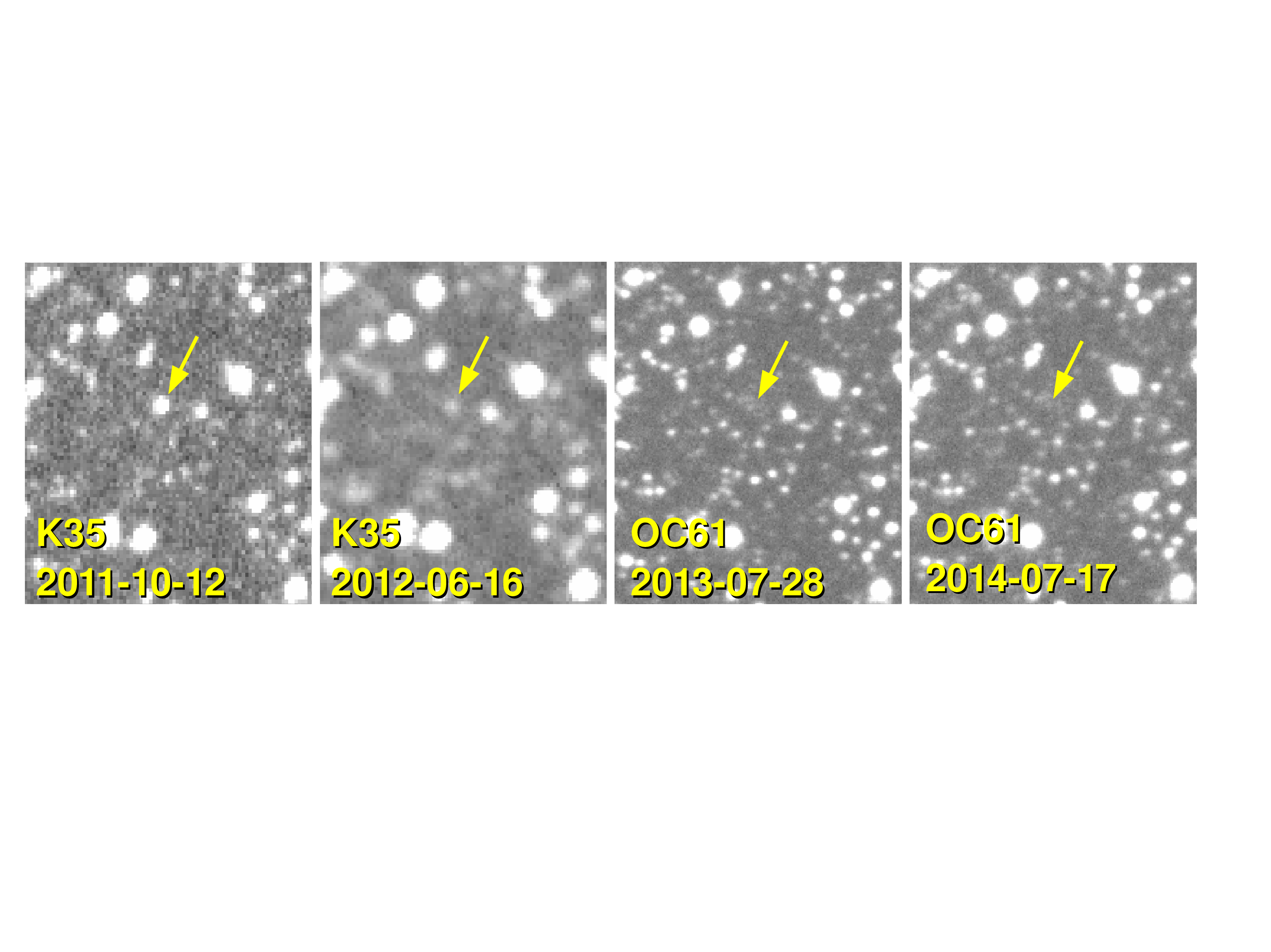}
\caption{Sequence of $V$-band images (1.7 arcmin on the long side,
North to the top, East to the left) of the advanced decline of
V5588 Sgr to pin-point the position of the nova and aid with the
identification of the progenitor and/or relic. The nova is seen
at $V$$\sim$20.5 on the 2013-07-28 image and it is fainter than
21.0 mag in the 2014-07-17 image.}
\end{figure}

At the nova position there is no counterpart on the Palomar I and II plates
or 2MASS and WISE catalogs.  At the reddening and distance of V5588 Sgr, a
M2-3 III cool giant similar to those present in novae RS Oph and T CrB would
shine at $K$$\sim$11.8, a sub-giant of the type present in nova U Sco at
$K$$\sim$15.2, and an M0 main sequence star at $K$$\sim$20.3. Considering that
the 2MASS survey did not detect stars fainter than 13.5 mag in $K$ within
2 arcmin of the position of the nova, we may only exclude a cool giant as
the donor star in V5588 Sgr, while a sub-giant and a main sequence donor
star are equally allowed by the 2MASS limiting magnitude. Similar
conclusions can be reached considering the $V$$\sim$21.1 upper limit to the
nova renmant brightness from our latest observation on July 17, 2014.

\section{Multiple secondary maxima}

The most striking feature displayed by V5588 Sgr is undoubtedly the six
secondary maxima (SdM for short) that adorn the lightcurve in Figure~1, with
Figure~5 providing a zooming on them all.  Their basic properties are
summarized in Table~5.

The SdM look similar, although not identical.  The time interval
between them increased along the series, with an interval of 18 days between
the first two and twice as much (42 days) between the last two.  Similarly,
the duration of SdM also increased along the series, from FWHM=2.1
days for the first to FWHM=8.6 days for the last.  The photometric
colors at peak SdM brightness changed along the six episodes.  Compared to
the underlying normal decline, the first three SdM were redder by 0.5 mag in
$B$$-$$V$ and bluer by the same mount in $V$$-$$I_{\rm C}$, indicating that
the energy distribution of the continuum source associated to the SdM was
sharply peaking at $V$-band wavelengths.  The color change associated with
the fourth SdM was similarly directed but of lower amplitude, while the last
two SdM were not accompanied by significant color changes. In the near-IR,
a minimal blueing in $J$$-$$K$ color was observed during the second SdM,
while the third saw no changes at all (cf.  Figure~3).

The first two SdM were brighter than the primary maximum of the nova
in all $B$$V$$R_{\rm C}$$I_{\rm C}$ bands, and the third equal in
$B$$V$$I_{\rm C}$ bands.  As mentioned above, the $\sim$2 days time delay
between peak brightness at optical and near-IR wavelengths (cf.  Figure~3)
suggests that broad-band emission during an SdM was governed by photospheric
expansion as in a fireball.  The projected areas of the expanding fireballs
associated with the first two SdM were therefore larger than the one for the
primary maximum.  The very rapid evolution of SdM in comparison with the
underlying normal nova evolution, suggests that a much lower amount of
material was ejected at a much larger velocity than at primary maximum.  The
fact that the emission during SdM was primarily continuum radiation is
confirmed by Figure~1, where the $R_{\rm C}$ lightcurve shows much less
pronounced SdM and the H$\alpha$ flux evolution does not follow the SdM
pattern.

We have built SdM profiles for $B$$R_{\rm C}$$I_{\rm C}$ similar to those
presented in Figure~5 for the $V$ band, and we have integrated their net
flux (i.e. the energy radiated in {\it excess} of the underlying normal nova
lightcurve). The sum of the energy radiated during SdM in the four optical
bands is listed in the last raw of Table~5 and plotted in Figure~6, where a
nice exponential decline is evident.

\begin{figure}
\includegraphics[width=84mm,height=21.5cm]{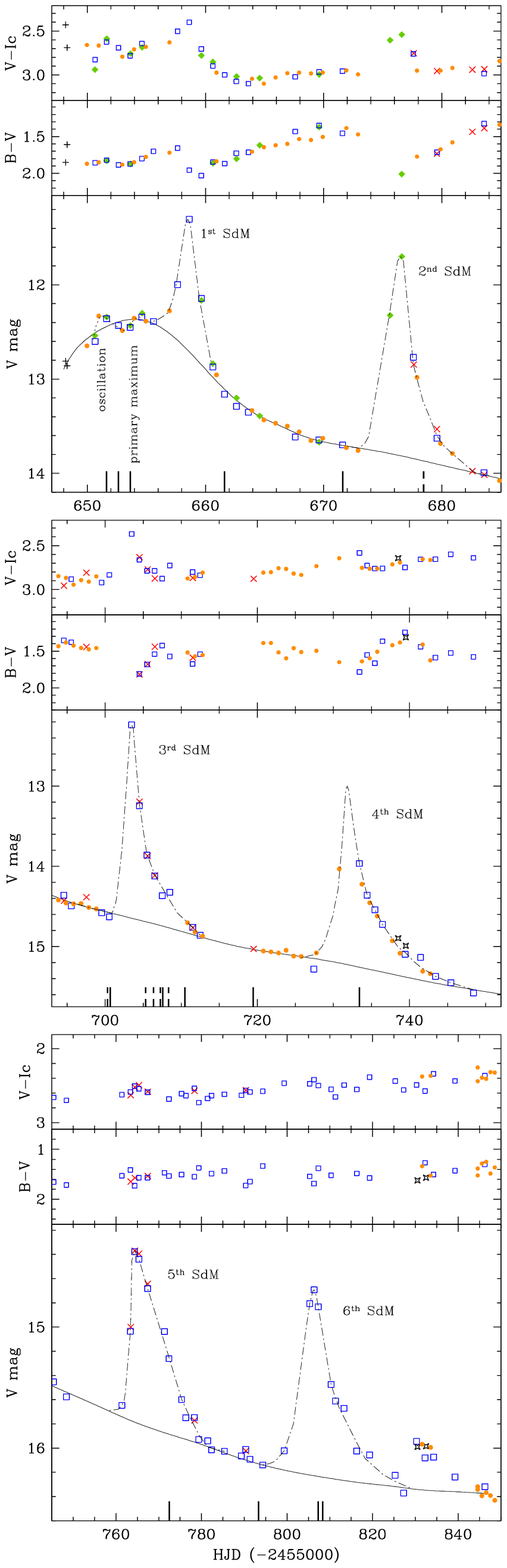}
\caption{Expanded view of the six secondary minima (SdM), highlighted by the
hand drawn dot-dashed lines.  The solid line represents the underlying normal
nova evolution.  The symbols for the different telescopes are identified in
Figure~1. Solid and dashed vertical bars mark the epochs of optical and
near-IR spectroscopic observations, respectively.}
\end{figure}

\begin{table}
 \centering
\caption{Basic parameters for the secondary maxima (SdM). The heliocentric
JD is HDJ + 2455000; $\Delta$t is the time elapsed between successive SdM;
FWHM measures the width in time of SdM; peak $V$ mag is listed next; $\Delta
V$ is the amplitude of SdM measured from the extrapolated underlying normal
nova decline (cf.  Figure~2); $M_V$ is the absolute magnitude reached by the
SdM after removing the contribution by the extrapolated underlying normal
nova decline; Flux$^{BVRI}_{bol}$ is expressed in
units of 10$^{-4}$ erg cm$^{-2}$ and represent the sum over the optical bands
of the net flux recorded during the SdM (i.e.  net the extra flux
superimposed to the underlying smooth nova decline), corrected for
$E_{B-V}$=1.56 reddening and a standard $R_V$=3.1 reddening law.}
\includegraphics[width=84mm]{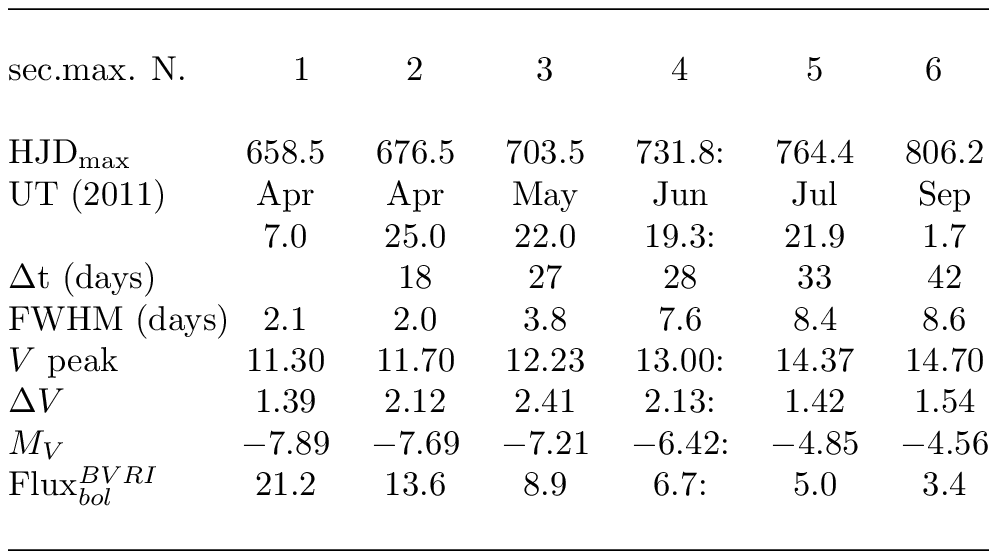}
\end{table}

\section{Spectroscopic evolution}

The spectral evolution at optical wavelengths is presented in Figure~7,
and that in the near-IR in Figures 9, 10 and 11, while Figure~8 highlights
the evolution of the profile of hydrogen emission lines.

\subsection{An hybrid nova}

At the time of discovery, the first confirmatory spectroscopic observation
by Arai et al.  (2011, on day $-$4.78) describes the spectrum of V5588 Sgr
as that of a typical FeII-type nova (Williams 1992) with prominent emission
lines of H$\alpha$ (FWHM=900 km/s), H$\beta$ and Fe II (multiplets 42, 48,
49) on a highly reddened continuum.  Our early optical spectra in Figure~7
(epochs $-$2.4 and $-$0.4 days) confirm the FeII classification, and the high
resolution H$\alpha$ profile for day +7.6 in Figure~8 shows a strong P-Cyg
absorption component (blue-shifted by 650 km/s with respect to the emission
component; FWHM(em)=770 km/s, FWHM(abs)=200 km/s) which is typical of FeII
novae around maximum light (e.g. McLaughlin 1960).

\begin{figure}
\includegraphics[width=84mm]{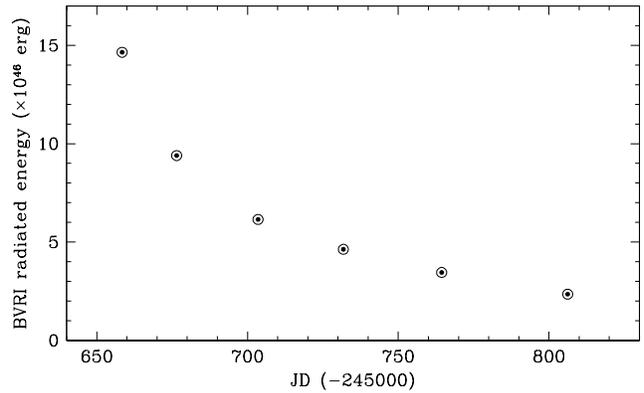}
\caption{Net energy radiated through $B$$V$$R_{\rm C}$$I_{\rm C}$ bands by
the six secondary maxima (for a distance to V5588 Sgr of 7.6 kpc, and
corrected for $E_{B-V}$=1.56 and $R_V$=3.1 extinction).}
\end{figure}

Our next spectrum in Figure~7, obtained on day +18, just before the rise to
second SdM, shows something quite unexpected: in addition to the initial
FeII-type spectrum, there is now simultaneously present the spectrum of a
fully flagged He/N-type nova, with a stunning diplay of HeI and HeII lines. 
On subsequent epochs the FeII-type spectrum declined in strength,
without developing the nebular lines (most notably [OIII]) that usually
dominates the advanced decline of FeII-type novae.  This agrees with the
fact that the lightcurves in Figure~1 do not show the typical flattening of
the brightness decline.  In FeII-type novae, the transition from optically
thick (emitting mostly permitted lines) to optically thin ejecta (dominated
by nebular lines) occurs between 3 and 4 mag below maximum, when the
photometric decline all of a sudden changes from rapid to a much flatter one
(McLaughlin 1960, Munari 2012).  While during the optically thick decline
the gas in the ejecta is just mainly recombining, during the optically thin
phase the hard radiation field from the central WD permeates the ejecta with
a strong re-ionization action.  There is available on the
web\footnote{http://kanatatmp.g.hatena.ne.jp/kanataobslog/20110416/p1}, a
spectrum for April 12 obtained with the Kanata 1.5m telescope that shows how
He/N-type spectrum was already emerging at that time (day +10), when V5588
Sgr had just ended the first SdM and was back to the smooth underlying
decline (see Figure~5).  We may therefore conclude that the appearance of
the He/N-type spectrum coincided with the nova entering the evolutionary
phase characterized by the SdM.  The last of our spectra still showing
(feeble) traces of the FeII-type spectrum is that for day +118, after that
date the recorded spectrum is only that of a He/N nova.

There are only very few novae that have been seen to evolve from a FeII- to
an He/N-type spectrum or to have simultaneously shown them both.  Williams
(1992) called them {\em hybrid} novae and reviewed their basic
characteristics.  He observed how they ($a$) display unusually broad FeII
lines, and ($b$) tend to evolve like normal He/N novae once the transition
from the initial FeII- to the final He/N-type has been completed. V5588 Sgr
distinguish itself from the other hybrid novae on both these points.

First of all V5588 Sgr displayed {\em narrow} emission lines during the FeII
phase, certainly not larger than seen in normal FeII novae.  As shown in
Figures~7 and 8, the FWHM of Balmer and FeII emission lines kept lower than
1000 km/s throughout the whole FeII phase.  In hybrid novae, the FWHM of
emission lines during the FeII phase is at least twice as large as seen in
V5588 Sgr, and remain similarly broad during the subsequent He/N phase.
Secondly, V5588 Sgr evolved slowly for a He/N. The $t_3$ characteristic
times of He/N novae are short, from a few days to a maximum of a few weeks,
while for V5588 Sgr it was a few months (2.5 months in $V$ and 3 in $B$
band, see Eq.(2) above). In addition, the width at half maximum of emission
lines in He/N novae is usually equal or larger than 2500 km/sec, and their
profile is more fat-topped than Gaussian-like. In V5588 Sgr the HeI and HeII
lines are stable at FWHM$\sim$1100 km/sec through out the recorded evolution,
with a profile well fitted by a Gaussian, marginally double-peaked with a
velocity separation of $\sim$400 km/sec. Finally during the advanced decline
of He/N novae, the intensity of HeII emission line is usually larger than
H$\beta$. In V5588 Sgr, HeII never grew in intensity to more than half of
H$\beta$.

The hybrid classification from optical spectra is not in contrast with
available observations in the near-IR.  Our first $JHK$ spectra were
obtained on day +24.5 (when optical spectra were already dominated by He/N
features and the signatures from the FeII phase were weakening), and their
appearance is typical of He/N novae, as are all the other near-IR spectra we
obtained (Figures 9, 10 and 11).  On the other hand, Rudy et al.  (2011)
labelled their near-IR spectrum of V5588 Sgr for day +25.5 as that of a FeII
nova, without the carbon lines.  The absence of carbon lines is
obvious in our spectra.  As shown in Banerjee and Ashok (2012), the FeII
class of novae early after outburst, are distinguished from the He/N class
by displaying a large number of strong carbon lines in each of the $J$,$H$
and $K$ bands.  Most prominent among these CI lines are the 1.165, 1.175
$\mu$m region features in the $J$ band and the strong cluster of lines
beyond Br10 in the $H$ band in the 1.74 to 1.80 $\mu$m region.  Examples of
these CI lines can be seen in the spectra of several FeII type novae viz.,
V1280 Sco and V2615 Oph (Das et al.  2008; Das, Banerjee $\&$ Ashok 2009),
V2274 Cyg (Rudy et al.  2003), V1419 Aql (Lynch et al.  1995) and V5579 Sgr
(Raj, Ashok $\&$ Banerjee 2011).  Earlier near-IR spectra, obtained around
primary maximum, would have been highly valuable to check if the FeII type
characterizing optical wavelengths at that time was similarly dominating the
near-IR.

\subsection{Coronal lines}

As noted by Williams (1992), the advanced evolution of He/N-type novae
proceeds in one of three distinct ways: ($a$) permitted emission lines
simply fade away and nebular emission lines are not observed, ($b$) strong
[NeIII] 3869, 3868 \AA\ o [NeV] 3346, 3426 \AA\ lines emerge, resulting in a
{\em neon} nova, or ($c$) coronal forbidden lines such as [FeX] 6375 \AA\
develop.

\begin{figure*}
\includegraphics[angle=270,width=175mm]{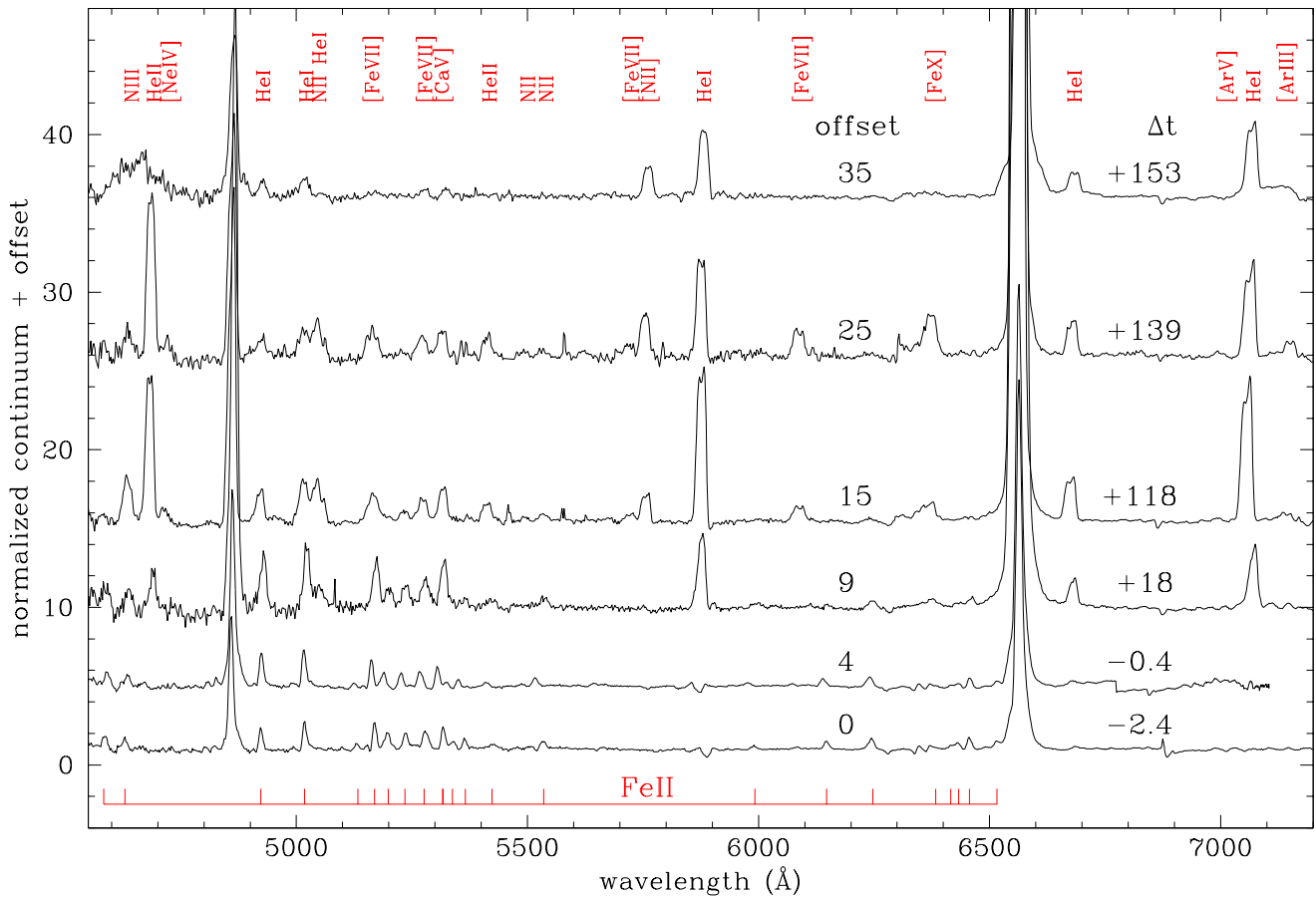}
\caption{Evolution of optical spectra, clearly showing the hybrid nature of
V5588 Sgr.}
\end{figure*}

V5588 Sgr did not show usual nebular lines. It also did not develop [NeV],
and only feeble traces of [NeIII] and [NeIV] were visible, therefore it did
not turn into a neon nova.  It instead developed prominent [FeX], stronger
than [FeVII] 6087 \AA\ at peak intensity on day +139 spectrum in Figure~7,
fully qualifying V5588 Sgr as a {\em coronal line} nova.  [ArX] 5534 \AA\
was probably present too, and [FeXI] 3987 could marginally be so.  [ArXI]
6019, [NiXV] 6702, [NiXIII] 5115 \AA\ which were prominent in nova RS Oph
(Wallerstein \& Garnavich 1986, Iijima 2009) were not present, while
not much can be said for [FeXIV] 5303 \AA\ given the presence of nearby and
strong [CaV] 5309 \AA.

[FeX] became visible for the first time on our spectrum for day +18, shortly
before the onset of the second SdM, simultaneous with the first appearance
of HeII in emission, signalling the ejecta were turning optically
transparent and nuclear burning was proceeding in the WD envelope. Under
such conditions, novae are usually detected as strong super-soft X-ray
sources (Krautter 2008). Unfortunately, the Swift satellite did not
observe V5588 Sgr until much later. The intensity of [FeX] increased
with time and in pace with HeII (cd Figure~7), peaking in intensity on our
spectrum for day +139 (2011 Aug 19), obtained during the normal decline
between fifth and sixth SdM. Quite interestingly, on our last spectrum,
obtained two weeks later (day +153) and right at the peak of sixth secondary
maximum, both HeII and [FeX] are instead missing. We see two alternative
explanation for this: either ($a$) the nuclear burning on the surface of the
WD stopped sometime between day +139 and +153 and the consequent rapid
cooling brought the temperature of the WD below the threshold for producing
HeII and [FeX], or ($b$) the material ejected during the sixth SdM
temporarily obscured the view of the WD and inner regions of the ejecta,
where the HeII and [FeX] form, or from where it was coming the hard
radiation capable of forming them in the external ejecta. The time scale for
recombination in the ejecta goes as:
\begin{equation}
t_{\rm rec} \approx 0.66 \left(\frac{T_{\rm e}}{10^4 {\rm ~K}}\right)^{0.8}
\left(\frac{n_{\rm e}}{10^9 {\rm ~cm}^{-3}}\right)^{-1} {\rm (hours)}
\end{equation}
following Ferland (2003). At the critical density for [FeX] (logN$_{\rm
crit}$=9.7 cm$^{-3}$), the recombination time scale is very short, less than
one hour for any reasonable choice of the electronic temperature $T_{\rm
e}$. Of course, after the switch-off of the nuclear burning, the envelope of
the WD is still very hot and cools off gradually, so the input of high
energy photons to the ejecta is not instantaneously stopped. Nonetheless,
there seem to be enough time between the spectra of +139 and +153 to allow
enough cooling of the WD to stop producing [FeX] lines from regions with an
electron density close to the critical value.

However, [FeVII] 6087 \AA\ too disappeared from day +153 spectrum. This line
came from regions characterized by an electron density lower that its
critical value (logN$_{\rm crit}$=7.6 cm$^{-3}$) but higher than that of
absent [OIII] lines (logN$_{\rm crit}$=5.8 cm$^{-3}$), so with a
recombination time from a few hours up to a couple of months.  The
disappearance of [FeVII] 6087 \AA\ line could therefore pose a problem for
alternative $"a"$ above. An even more serious problem comes from the text of
an approved Target of Opportunity proposal (available on the web) for Swift
X-ray observations of V5588 Sgr to be carried out in 2012 that states the
nova was displaying at the time of proposal submission (2012 May 18, or day
+411) a [FeX] emission line half the intensity of [FeVII] on optical
spectra, and thus observations were requested to observe the expected
super-soft X-ray emission.

\begin{figure*}
\includegraphics[width=175mm]{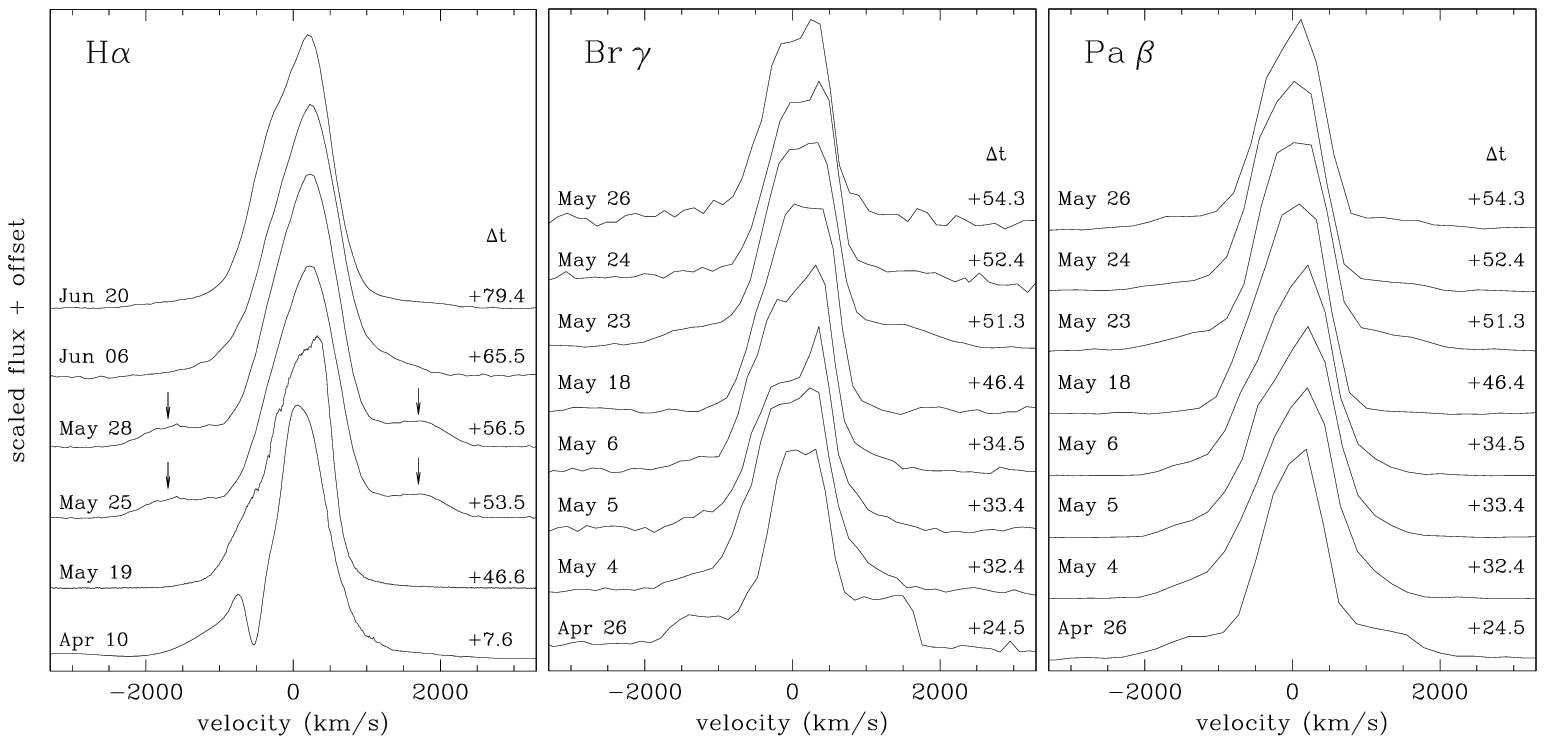}
\caption{Profiles at different epochs of hydrogen H$\alpha$, Br$\gamma$ and
Pa$\alpha$ emission line profiles. The arrows on the left panel highlight
the presence of the broad pedestal appearing at selected dates.}
\end{figure*}

We are tempted to conclude that [FeX] was still visible long after the end
of our optical spectroscopic monitoring, and that the absence of [FeX] and
HeII from the day +153 spectrum at the peak of the sixth SdM was caused by
optically thick material ejected during such SdM blocking the view toward
inner ejecta and central star. Unfortunately, for none of the other SdM
there are optical spectra obtained during the rise in brightness or around
maximum to check if this was a repeating pattern through all observed SdM.

\subsection{The variable two-components line profiles}

The emission lines of V5588 Sgr displayed throughout the whole recorded
outburst show a narrow profile, with FWHM$\sim$1000 km/sec for both
optical and near-IR spectra.  However, sometimes a much broader and weak
pedestal appeared at the bottom of the narrow and much stronger component. 
This is well seen in Figure~8, where the broad pedestal in obvious in the
H$\alpha$ profiles for days +53.5 and +56.5, or the Br$\gamma$ and Pa$\beta$
profiles for days +24.5 and +51.3.  On the H$\alpha$ profiles of Figure~8,
the pedestal has a trapezoidal shape, extending for $\Delta$vel=3600 km/sec
at the top and 4500 km/sec at the bottom.  The shape is similar for
Br$\gamma$ and Pa$\beta$, with reduced velocities: 2900 at the top and 3600
at the bottom.  The limited resolution of our spectra does not allow to
distinguish between a boxy, flat-topped profile and two separate emission
symmetric with respect to the main component, in other words if the emission
from the pedestal originates in a filled prolate volume or in a bipolar
arrangement.

A most important fact to note is that, both at optical and near-IR
wavelengths, the broad pedestal to Hydrogen lines is not seen in between
SdM, or during their rising toward maximum and early decline from it, but
only during the advanced decline from SdM maximum and before the return to
the smooth decline between SdM. The broad component developed similarly for
HeI lines, as Figure~11 well illustrates from $K$-band spectra: on Apr 26
and May 23, when the pedestal was present in Br$\gamma$, it was so also for
HeI 2.0581 $\mu$m, while on other dates only the narrow component was
visible in HeI as for Br$\gamma$. The pedestal to HeI is prominent in
near-IR spectra, while it is too faint to be seen in optical spectra.

\begin{figure}
\includegraphics[width=84mm]{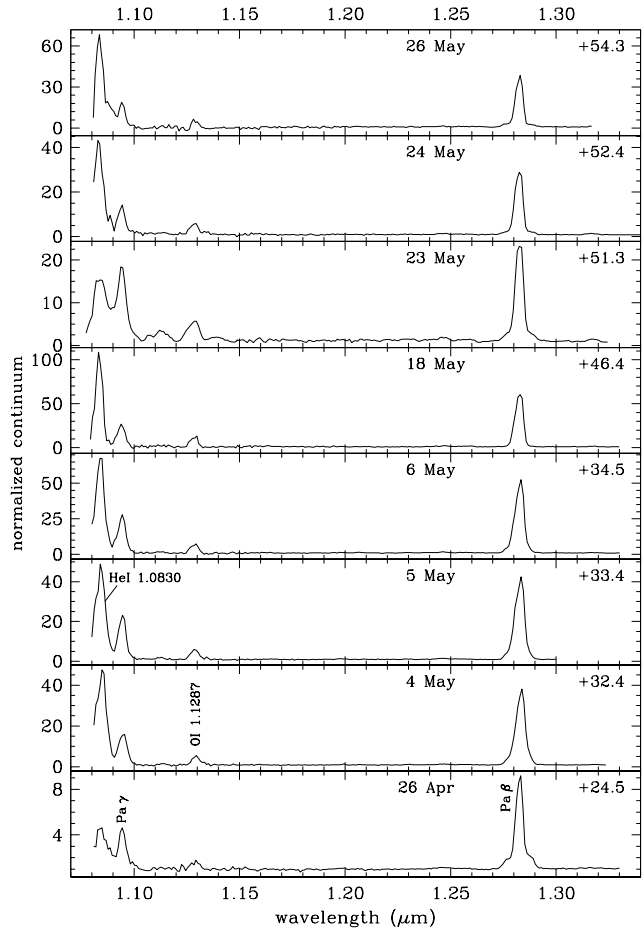}
\caption{$J$ band spectra of V5588 Sgr with prominent lines marked. Weak
lines of OI 1.3164 $\mu$m and NI 1.2461,69 $\mu$m are also detected which
can be seen if the spectra are magnified.  All spectra are normalized to
unity at 1.25 $\mu$m at which the corresponding observed fluxes are (in
chronological order) 11.8, 8.7, 8.44, 8.13, 4.64, 7.62, 6.83, 6.11 and
5.47$\times$$10^{-17}$ W/$\rm cm^{2}$/$\mu$m.}
\end{figure}

\begin{figure}
\includegraphics[width=84mm]{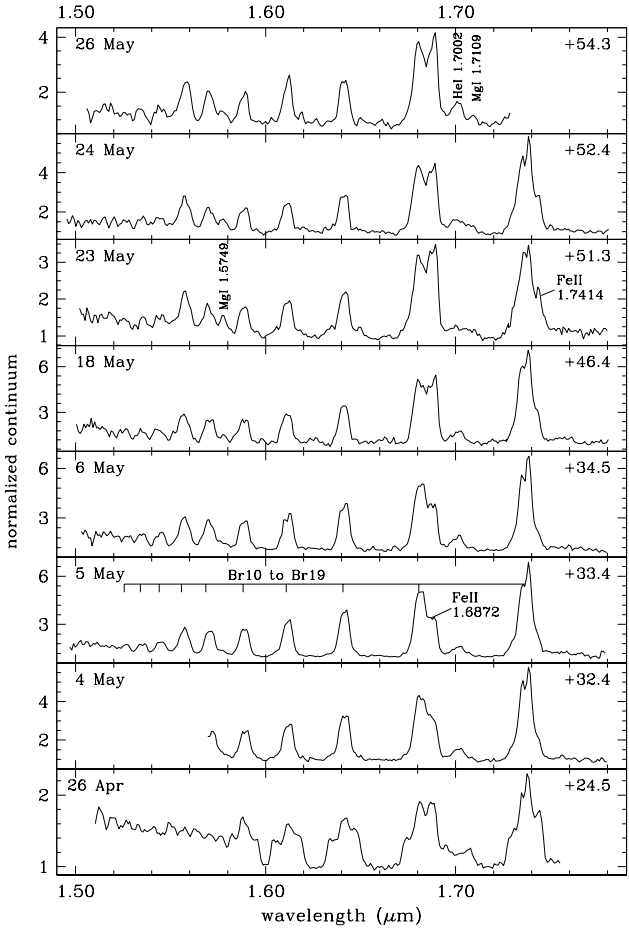}
\caption{The $H$ band spectra of V5588 Sgr. All spectra are normalized to
unity at 1.65 $\mu$m at which the corresponding observed fluxes are (in
chronological order) 9.75, 3.7, 3.47, 3.26, 2.0, 3.8, 3.29, 2.81,
2.43$\times$$10^{-17}$ W/$\rm cm^{2}$/$\mu$m.}
\end{figure}

\begin{figure}
\includegraphics[width=84mm]{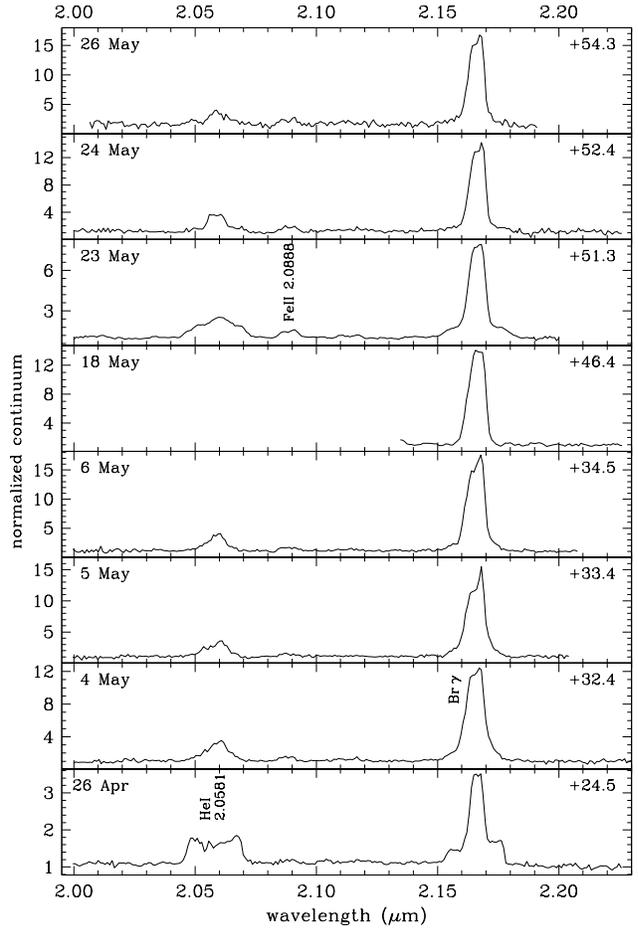}
\caption{$K$ band spectra of V5588 Sgr with prominent lines marked. All
spectra are normalized to unity at 2.2 $\mu$m at which the corresponding
observed fluxes are (in chronological order) 6.7, 2.8, 2.71, 2.64, 1.45,
2.62, 2.41, 2.2, 2.02$\times$$10^{-17}$ W/$\rm cm^{2}$/$\mu$m.}
\end{figure}

\subsection{A few NIR lines worth a special mention}

While most of the HI, HeI and OI lines in Figures 9, 10 and 11  are
routinely seen in the NIR spectra of novae (Banerjee and Ashok, 2012), there
are a few features which need special mention. These are the features at
1.6872 and 1.7414 $\mu$m in the $H$ band and a $K$ band feature at $\sim$
2.089 $\mu$m. Both the $H$ band features are clearly seen in Figure 3 and
although their origin is not certain, they could possibly be due to Fe II. 
The 1.6872 $\mu$m line is seen to slowly gain in strength to equal and even
surpass the adjacent Br11 line.  In the near-infrared, there are a few FeII
lines seen in the spectra of novae, which are believed to be primarily
excited by Lyman $\alpha$ and Lyman continuum fluorescence.  Among these are
the so-called "one micron Fe II lines" seen at around the 1 micron region in
several novae (Rudy et al. 2000 and references therein).  In addition, two
other Fe II lines at 1.6872 and 1.7414 $\mu$m in the $H$ band, have also
proposed to be pumped by the same mechanism (Banerjee, Das $\&$ Ashok 2009;
Bautista et al.  2004).  The $H$ band lines are prominently detected in the
2006 outburst of recurrent nova RS Oph (Banerjee, Das $\&$ Ashok 2009), in
the slow nova V2540 Oph (Rudy et al 2002a), in V574 Pup (Naik et al.  2010)
and possibly also in the recurrent nova CI Aql (Lynch et al.  2004).  As the
detections of these lines in individual objects increase, it becomes amply
evident that these $H$ band lines could be present in the spectra of other
novae too, but have evaded detection because of blending - especially when
their widths are large - with the Br 11 (1.6806 $\mu$m) line.

An emission feature  at $\sim$2.090 $\mu$m in the $K$ band is also seen in
the spectra at most of the epochs of observation.  Two possible
identifications may be considered for this line.  First, it could be FeII
2.0888 $\mu$m, a line also seen in the nova V2615 Oph (Das, Banerjee $\&$
Ashok 2009)for which an excitation mechanism by Lyman $\alpha$ fluorescence
was proposed.  Alternatively, this feature could be the [Mn XIV] 2.0894
$\mu$m coronal line which has been seen in a few instances in novae spectra
during the coronal phase viz., in nova V1974 Cyg (Wagner $\&$ Depoy, 1996)
and in RS Oph (Banerjee, Das $\&$ Ashok 2009).  It should be noted that the
$\sim$2.090 $\mu$m line should not be confused with an unidentified
line at 2.0996 $\mu$m that has often been detected in novae and which still
remains unidentified (Rudy et al.  2002b).

\section{Discussion}

Given the wide range of intriguing and unique features presented by V5588
Sgr, it was unfortunate that the nova was so distant and suffered extinction
to such an extent that it was rather faint at maximum. This inhibited
raising widespread attention among observers and thereby allowing larger
telescopes to provide higher resolution and more abundant spectral
observations.

The most fascinating aspect of  V5588 Sgr is undoubtedly the repeated
presence of nearly identical secondary maxima.  The body of evidence
presented in this paper suggests that they resulted from ejection at
high velocity of a limited amount of material.  The rise toward SdM maximum
corresponds to the initial expansion in optically thick conditions, the
maximum brightness to the maximum projected area reached by the expanding
pseudo-photosphere, and the decline by the material turning optically thin
and dissolving into surrounding space.

The extremely fast evolution of the SdM argues in favor of a limited amount
of material being ejected during such episodes, with the rise to maximum
taking generally less than one day and only a few days to complete the
decline.  It may also be argued that the ejection was in the form of a thin
and almost empty shell.  In fact, while its optically thick
pseudo-photosphere attained a maximum a brightness in excess of that of the
primary nova ejecta, as soon as it turned optically thin, its
continuum emission dropped like a stone and the only emission line feature
detectable was in the form of the weak pedestal, briefly seen in Hydrogen
and HeI lines.

It seems that the high velocity material ejected during SdM did not
collide significantly with the pre-existing slow velocity normal nova
ejecta.  There are no obvious traces of shocked or decelerating gas in the
optical and near-IR spectra.  In addition, the Swift X-ray satellite
observed V5588 Sgr seven times between 30 August 2011 (while the nova was on
the rise toward its sixth SdM) and 24 July 2012.  The nova was only detected
during the 2011 observations, at a very low average count rate of of 0.010
+/- 0.002 c/s, in contrast with the expectations from high velocity SdM
material slamming onto the normal nova ejecta.  Therefore, the material
ejected during SdM either expanded and dissolved into the cavity left over
by expanding normal ejecta or moved along spatial directions away from
spatially confined main ejecta (for ex.  with the normal ejecta expanding on
the plane of the sky in a bipolar shape, and the high velocity SdM material
arranged orthogonal to this like in an equatorial ring or a spherical
central blob).

Only the latest theoretical models of thermonuclear runaways on novae
(e.g.  Hillmann et al.  2014) are able to produce the type of secondary
maxima seen in V5588 Sgr.  Their niche in the parameter space will need
further exploration, and the extensive and accurate observational data here
provided for V5588 Sgr will surely help in fine tuning the models.  External
sources not usually accounted for by models on thermonuclear runaways on
white dwarfs may also play a role, like for example the donor star being
able to refuel the WD with short and repeated pulses of mass transfer. 
Whatever the reason was, it is striking that the energy release during SdM
(at least the energy radiated through the optical bands, cf.  Figure~6)
declined following an exponential pattern, while at the same time the breath
of the SdM was increasing as it was the time interval between two successive
ones, a pattern suggesting a relaxation mechanism.  In the case of the donor
star refueling the WD with pulses of mass transfer, the exponential decline
in energy radiated by SdM suggests that the amount of transferred mass
followed a similar decline.  Such mass transfer episodes does not seem being
triggered by orbital geometry (like passages at periastron in a highly
eccentric and long period orbit), because of the monotone increase in time
separation between them poorly compares with the regularity of orbital
revolutions.

The EVLA radio observations summarized by Krauss et al. (2011c) and the
comparison with V4745 Sgr (Nova 2003 N.1), the only other known nova with a
lightcurve resembling V5588 Sgr, does not point to an easy solution. In
fact, V5588 Sgr was found to be radio quiet just before SdM N.2 and radio
loud just before SdM N. 3 and 4, and it was radio quiet during SdM N.5 and
immediately after the end of SdM N.2 and N.3. As noted by Krauss et al.
(2011c) such rapid and repeated on-off switch in radio emission cannot be
easily understood in the conventional modeling for radio emission of nova
ejecta. Both V5588 Sgr and V4745 Sgr displayed several SdM, over a similar
time interval and of similar brightness. However, V4745 Sgr begun as a FeII
nova and did not developed a He/N spectrum at later times, e.g. it was not
an hybrid nova, and its emission lines were twice as wide as in V5588 Sgr.
In addition, while the spectra of V5588 Sgr did not change during SdM other
than for the appearance of the weak broad pedestal, the emission lines of
V4745 Sgr during SdM switched back to strong P-Cyg absorption profiles as
for the primary maximum (Cs{\'a}k et al. 2005, Tanaka et al. 2011).

A dedicated comparison of detailed properties of the few novae that
displayed one or more SdM is well beyond the scope of this paper, but surely
worth considering. While information like orbital period and orbital
inclination or presence of WD modulation (suggesting a significant magnetic
field) are available for some of these novae, the same will be hardly
achievable for V5588 Sgr given its faint magnitude in quiescence ($V$$>$21
mag), requiring the largest available telescopes for the attempt.

\section{Acknowledgements} We would like to thank E. Tamajo, P. Valisa and
P. Ochner for assistance with some of the spectroscopic observations. The
research work at the Physical Research Laboratory is funded by the
Department of Space, Government of India.


\begin{thebibliography}{}
\bibitem[\protect\citeauthoryear{Arai}{2011}]{b10}Arai A., Nagashima M., Kajikawa T., and Naka C., 2011, IAUC 9203
\bibitem[\protect\citeauthoryear{banerjee}{2002}]{b13} Banerjee D.P.K. $\&$ Ashok, N.M., 2002,A$\&$A, 395, 161
\bibitem[\protect\citeauthoryear{banerjee}{2012}]{b73} Banerjee D.P.K. $\&$ Ashok, N.M., 2012, BASI, 40, 243    
\bibitem[\protect\citeauthoryear{Banerjee}{2009}]{b10} Banerjee D.P.K., Das R.K., Ashok N.M., MNRAS, 2009, 399, 357
\bibitem[\protect\citeauthoryear{Banerjee}{2011}]{b10} Banerjee D.P.K. $\&$ Ashok N.M., 2011, Astronomers Telegram, 3345
\bibitem[\protect\citeauthoryear{Bautista}{2004}]{b4} Bautista M. A., Rudy R. J., Venturini C. C., 2004, ApJ, 604, L129
\bibitem[\protect\citeauthoryear{Buscombe \& de Vaucouleurs}{1955}]{1955Obs....75..170B} Buscombe W., de Vaucouleurs G., 1955, Obs, 75, 170 
\bibitem[\protect\citeauthoryear{Capaccioli et al.}{1989}]{1989AJ.....97.1622C} Capaccioli M., della Valle M., Rosino L., D'Onofrio M., 1989, AJ, 97, 1622 
\bibitem[\protect\citeauthoryear{Cs{\'a}k et al.}{2005}]{2005A&A...429..599C} Cs{\'a}k B., Kiss L.~L., Retter A., Jacob A., Kaspi S., 2005, A\&A, 429, 599
\bibitem[\protect\citeauthoryear{Das}{2008}]{b2} Das R.K., Banerjee D.P.K., Ashok N.M.,Chesneau O., 2008, MNRAS, 391, 1874
\bibitem[\protect\citeauthoryear{Das}{2009}]{b4} Das R.K.,  Banerjee D.P.K., Ashok N.M., 2009, MNRAS, 398, 375
\bibitem[\protect\citeauthoryear{Downes \& Duerbeck}{2000}]{2000AJ....120.2007D} Downes R.~A., Duerbeck H.~W., 2000, AJ, 120, 2007
\bibitem[\protect\citeauthoryear{Ferland}{2003}]{2003ARA&A..41..517F} Ferland G.~J., 2003, ARA\&A, 41, 517
\bibitem[\protect\citeauthoryear{Fiorucci \& Munari}{2003}]{2003A&A...401..781F} Fiorucci M., Munari U., 2003, A\&A, 401, 781 
\bibitem[\protect\citeauthoryear{Harman}{2008}]{b20} Harman D. J. et al., 2009, ASPC, 401, 246, (astro-ph 0809.4592)
\bibitem[\protect\citeauthoryear{Hounsell et al.}{2010}]{2010ApJ...724..480H} Hounsell R., et al., 2010, ApJ, 724, 480 
\bibitem[\protect\citeauthoryear{Iijima}{2009}]{2009A&A...505..287I} Iijima T., 2009, A\&A, 505, 287 
\bibitem[\protect\citeauthoryear{Kiyota}{2011}]{2011IAUC.9203....1K} Kiyota S., 2011, IAUC, 9203, 1
\bibitem[\protect\citeauthoryear{Krauss et al.}{2011}]{2011ATel.3319....1K} Krauss M.~I., et al., 2011a, ATel, 3319
\bibitem[\protect\citeauthoryear{Krauss et al.}{2011}]{2011ATel.3397....1K} Krauss M.~I., et al., 2011b, ATel, 3397
\bibitem[\protect\citeauthoryear{Krauss et al.}{2011}]{2011ATel.3539....1K} Krauss M.~I., et al., 2011c, ATel, 3539
\bibitem[\protect\citeauthoryear{Krautter}{2008}]{2008ASPC..401..139K} Krautter J., 2008, ASPC, 401, 139 
\bibitem[\protect\citeauthoryear{Landolt}{2009}]{2009AJ....137.4186L} Landolt A.~U., 2009, AJ, 137, 4186 
\bibitem[\protect\citeauthoryear{Lynch}{1995}]{b23} Lynch D.K., Rossano G.S., Rudy R.J., Puetter R.C, 1995, AJ, 110, 2274
\bibitem[\protect\citeauthoryear{Lynch}{2004}]{b26} Lynch D. K., Wilson J. C., Rudy R. J., Venturini C., Mazuk S., Miller N.A., Puetter R. C., 2004, AJ, 127, 1089
\bibitem[\protect\citeauthoryear{McLaughlin}{1960}]{1960stat.conf..585M} McLaughlin D.~B., 1960, in Stellar Atmospheres. J. L. Greenstein ed., University of Chicago Press, 585 
\bibitem[\protect\citeauthoryear{Maehara}{2011}]{2011IAUC.9203....1M} Maehara H., 2011, IAUC, 9203, 1
\bibitem[\protect\citeauthoryear{Munari et al.}{2008a}]{2008A&A...492..145M} Munari U., et al., 2008a, A\&A, 492, 145
\bibitem[\protect\citeauthoryear{Munari et al.}{2008b}]{2008MNRAS.387..344M} Munari U., Henden A., Valentini M., Siviero A., Dallaporta S., Ochner P., Tomasoni S., 2008, MNRAS, 387, 344
\bibitem[\protect\citeauthoryear{Munari et al.}{2010}]{2010PASP..122..898M} Munari U., Henden A., Valisa P., Dallaporta S., Righetti G.~L., 2010, PASP, 122, 898 
\bibitem[\protect\citeauthoryear{Munari}{2011a}]{b10} Munari U.,  et al. 2011a, CBET 2707
\bibitem[\protect\citeauthoryear{Munari}{2011b}]{b10} Munari U.,  Ashok N. M., Banerjee D. P. K., Righetti G. L., Dallaporta S., Cetrulo G., Englaro A., 2011b, CBET 2723
\bibitem[\protect\citeauthoryear{Munari et al.}{2011c}]{2011NewA...16..209M} Munari U., Siviero A., Dallaporta S., Cherini G., Valisa P., Tomasella L., 2011c, NewA, 16, 209
\bibitem[\protect\citeauthoryear{Munari et al.}{2012}]{2012BaltA..21...13M} Munari U., et al., 2012, BaltA, 21, 13 
\bibitem[\protect\citeauthoryear{Munari \& Moretti}{2012}]{2012BaltA..21...22M} Munari U., Moretti S., 2012, BaltA, 21, 22 
\bibitem[\protect\citeauthoryear{Munari}{2012}]{2012JAVSO..40..582M} Munari U., 2012, JAVSO, 40, 582 
\bibitem[\protect\citeauthoryear{Munari \& Valisa}{2014}]{2014CoSka..43..174M} Munari U., Valisa P., 2014, in Observing Techniques, Instrumentation and Science for Metre-Class Telescopes, T. Pribulla ed., CoSka, 43, 174 
\bibitem[\protect\citeauthoryear{Naik}{2010}] {b10} Naik S., Banerjee D. P. K., Ashok N. M., Das, R.K., 2010, MNRAS, 404, 367
\bibitem[\protect\citeauthoryear{Nishiyama et al.}{2011}]{2011IAUC.9203....1N} Nishiyama K., Kabashima F., 2011, IAUC, 9203, 1
\bibitem[\protect\citeauthoryear{Pejcha}{2009}]{2009ApJ...701.L119P} Pejcha, O. 2009, ApJ, 701, L119
\bibitem[\protect\citeauthoryear{Raj, Ashok, \& Banerjee}{2011}]{2011MNRAS.415.3455R} Raj A., Ashok N.~M., Banerjee D.~P.~K., 2011, MNRAS, 415, 3455 
\bibitem[\protect\citeauthoryear{Rudy}{2000}]{b35} Rudy R. J., Puetter R. C., Mazuk S., Hamann, F., 2000, ApJ, 539, 166
\bibitem[\protect\citeauthoryear{Rudy}{2002a}]{b36} Rudy R. J., Lynch D. K., Mazuk S., Venturini C. C., Puetter R. C., Perry R. B., 2002, BAAS, 34, 1162
\bibitem[\protect\citeauthoryear{Rudy}{2002b}]{b26}  Rudy R. J., Venturini C., Lynch D. K., Mazuk S.,  Puetter R. C., 2002, ApJ, 573, 794
\bibitem[\protect\citeauthoryear{Rudy}{2003}]{b34} Rudy R. J., Dimpel W. L., Lynch D. K., Mazuk S., Venturini C. C., Wilson J. C., Puetter R. C., Perry R. B., 2003, ApJ, 596, 1229
\bibitem[\protect\citeauthoryear{Rudy}{2011}]{b10}R. J. Rudy, R. W. Russell, M. Sitko, 2011, IAUC 9211
\bibitem[\protect\citeauthoryear{Seaquist \& Bode}{2008}]{SB}Seaquist, E. R., \& Bode, M. F. 2008, in Cambridge Astrophys. Ser. 43, Classical Novae, eds., M. F. Bode \& A. Evans, (2nd ed., Cambridge: Cambridge Univ. Press), 141
\bibitem[\protect\citeauthoryear{Strope et al.}{2010}]{2011AJ....140...34S} Strope R. J., Schaefer B. E., Henden A. A., 2010, AJ, 140, 34  
\bibitem[\protect\citeauthoryear{Tanaka et al.}{2011}]{2011PASJ...63..159T} Tanaka J., Nogami D., Fujii M., Ayani K., Kato T., 2011, PASJ, 63, 159
\bibitem[\protect\citeauthoryear{van den Bergh \& Younger}{1987}]{1987A&AS...70..125V} van den Bergh S., Younger P.~F., 1987, A\&AS, 70, 125 
\bibitem[\protect\citeauthoryear{Venturini at al.}{2004}]{vent} Venturini, C. C., Rudy, R. J., Lynch, D. K.,Mazuk, S., Puetter, R. C., et al. 2004, AJ, 128, 405
\bibitem[\protect\citeauthoryear{Wagner}{1996}]{b45} Wagner R. M., Depoy D. L., 1996, ApJ, 467, 860
\bibitem[\protect\citeauthoryear{Wallerstein \& Garnavich}{1986}]{1986PASP...98..875W} Wallerstein G., Garnavich P.~M., 1986, PASP, 98, 875 
\bibitem[\protect\citeauthoryear{Williams}{1992}]{b131} Williams R.E., 1992, AJ, 104, 725

\end{thebibliography}
\end{document}